\documentclass[aps,psfig,epsfig,preprint]
{revtex4}
\usepackage{epsfig}

\newcommand{\pa}{\partial}

\begin{document}
\draft
\title{
{\normalsize \hskip4.2in USTC-ICTS-05-03} \\{\bf Proton-Antiproton
Annihilation in Baryonium }}
\author{Gui-Jun Ding\footnote{e-mail address: dinggj@mail.ustc.edu.cn},
Mu-Lin Yan\footnote{e-mail address: mlyan@ustc.edu.cn; corresponding
author.}}

\affiliation{}

\address{
 Interdisciplinary Center for Theoretical Study, University
of Science and Technology of China, \\ Hefei, Anhui 230026, China}

\date{}
\begin{abstract}
A possible interpretation of the near-threshold enhancement
 in the $(p\bar{p})$-mass spectrum in $J/\psi{\to}\gamma p{\bar
p}$ is the of existence of a narrow baryonium resonance $X(1860)$.
Mesonic decays of the $(p\bar{p})$-bound state $X(1860)$ due to
the nucleon-antinucleon annihilation are investigated in this
paper. Mesonic coherent states with fixed $G$-parity and
$P$-parity have been constructed . The
Amado-Cannata-Dedoder-Locher-Shao formulation(Phys Rev Lett. {\bf
72}, 970 (1994)) is extended to  the decays of the $X(1860)$. By
this method, the  branch-fraction ratios of $Br(X\rightarrow \eta
4\pi)$, $Br(X\rightarrow \eta 2\pi)$ and $Br(X\rightarrow 3\eta )$
are calculated. It is shown that if the $X(1860)$ is a bound state
of $(p\bar{p})$, the  decay channel  ($X\rightarrow \eta 4\pi)$ is
favored over  $(X\rightarrow \eta 2\pi)$. In this way, we develop
 criteria for distinguishing the baryonium interpretation for the
near-threshold enhancement effects in $(p\bar{p})$-mass spectrum
in $J/\psi{\to}\gamma p{\bar p}$ from other possibilities.
Experimental checks are expected. An intuitive picture for our
results is discussed.
\end{abstract}


\maketitle

\section{Introduction}

\noindent There is growing interest in  exotic hadrons, which may
open new windows for understanding the hadronic structures and QCD
at low energy. Recently, the BES Collaboration observed a
near-threshold enhancement in the proton-antiproton $(p\bar{p})$
mass spectrum from the radiative decay $J/\psi\rightarrow \gamma
p\bar{p}$ \cite{BES}. This enhancement can be fitted with either
an $S$- or $P$-wave Breit-Wigner resonance function. In the case
of $S$-wave fit, the peak mass is at $M=1859_{-10}^{+3}({\rm
stat})_{-25}^{+5}({\rm sys})$ with a total width $\Gamma<
30~\mbox{MeV/c}^2$ at $90\%$ percent confidence level. For the
$S-$wave fit, the corresponding spin and parity are
$J^{PC}=0^{-+}$. Moreover, the Belle Collaboration also reported
similar observations of the decays $B^+\rightarrow K^+p\bar{p}$
\cite{Bell1} and $\bar{B}^0\rightarrow D^0p\bar{p}$ \cite{Bell2},
showing enhancements in the $p\bar{p}$ invariant mass
distributions near $2m_p$. These observations could be naively
interpreted as signals for baryonium $p\bar{p}$ bound states
\cite{Yang}\cite{Datta}\cite{Yan}. The BES-datum fit in
ref.\cite{BES}
 represents the simplest interpretation of the
experimental results as an indication of a baryonium resonance.
Here, we denote this baryonium particle as $X(1860)$ with
$J^{PC}=0^{-+}$ and $I^G=0^+$\cite{Peking}.
 However, this is only one possible interpretation. Other
possible ways to understand this phenomena include, for instance,
a flavorless gluon state\cite{Rosner}, a final state
interactions\cite{Zou} or an effect  of the quark fragmentation
process\cite{Rosner}, etc. In order to ascertain whether or not
the  $X(1860)$ exists, more evidence is needed. A significant
distinguishing feature for the baryonium interpretation is that
the decays of $X(1860)$ are mainly due to proton-antiproton
annihilation in the baryonium. In this paper, we investigate
$p\bar{p}$-annihilations in $X(1860)$ by means of a coherent-state
method on nucleon-antinucleon annihilation in large $N_c$
QCD\cite{batch1}\cite{batch2}.

At first glance, the most favorable $X(1860)$ decay channel would
be $X(1860)\rightarrow \eta \pi\pi$,  because it is the simplest
hadronic process with the largest phase space.  However, since the
decay is caused by  $(p\bar{p})$-annihilations in the $X(1860)$,
this naive observation may be not true. Low-energy
nucleon-antinucleon annihilation is a fertile area for studying
low energy QCD, and there are many experimental and theoretical
studies in the literature(e.g., see \cite{batch1}\cite{Yang
lu3}\cite{batch2}\cite{Sedlak}\cite{Amsler}
\cite{Sommermann}\cite{Amado97}\cite{Amado94}). In these reports,
it has been shown that the nucleon-antinucleon annihilation at
rest mostly favors processes with between 4 and 7 pion final
states, over those with two or three pion's \cite{Amado97}. This
is a a general characteristic of $(p\bar{p})$-annihilation
(without consideration of  $J^{PC}$ and $I^G$ quantum numbers). It
is interesting to pursue whether or not there are similar features
in $X(1860)$  decay. If so, we will have a new criteria to
characterize the $X(1860)$. This is the main aim of the work
reported here.

The contents of this paper are organized as follows: In section
II, we use a toy model to describe the possibility of the
$(p-\bar{p})$-collisions inside a ($p\bar{p}$)-bound state; in
section III, we construct the mesonic coherent states with fixed
$G$- and $P$-parities; Section IV describes calculation of the
branch fractions of $Br(X\rightarrow \eta 4\pi)$, $Br(X\rightarrow
\eta 2\pi)$ and $Br(X\rightarrow 3\eta )$. Finally, we briefly
summarize our results, and provide an intuitive picture for our
results.

\section{A toy model description on $(p-\bar{p})$ collisions
inside a ($p\bar{p}$)-bound state}

In order to understand the mechanics of $X(1860)$ decays, we {\it
assume} that  the proton and the antiproton collide with each
other  inside the $X(1860)$, resulting in the collapse of the
$(p\bar{p})$-bound state , or the decay due to the effect of rapid
nucleon-antinucleon annihilation. We use the
$(p-\bar{p})$-collision frequency $\nu$ (i.e., the collision times
per time-unit period) in the $X(1860)$ to characterize the
possibility of such $(p\bar{p})$-collisions. We show that this
frequency can be estimated in a simple toy model, in which the
proton and the antiproton are treated as point-like particles.
This collision frequency is actually equivalent to the total
$X(1860)$ decay width in the {\it annihilation-assumption}
mentioned above. Following ref.\cite{Yan},  we roughly  sketch the
toy model for the $X(1860)$  $(p\bar{p})$-bound state. The
single-well potential toy model for the $(pn)$-molecular bound
state deuteron first appeared in the literature  more than fifty
years ago \cite{Ma}. The quasi-stable $(p\bar{p})$-molecular bound
state $X(1860)$ can
 be similarly described by a double-well potential(or
Skyrmion-type potential) toy model\cite{Datta}\cite{Yan}. The
potential of such a double-well model, $V(\rho)$, is expressed as
follows\cite{Yan} (see Fig.1)
\begin{equation}\label{potential}
V(\rho)=2m_p-c\;\delta(\rho)+V_c(\rho),
\end{equation}
where
\begin{equation}
V_c(\rho)=\left\{
   \begin{array}{ll}
   {h={m_p/4}},\ \ & 0<\rho<\lambda,\\
   -V_{p\bar{p}}
   =-73~\mbox{MeV},     & \lambda<\rho<a_{p\bar{p}},\\
   0, &~~~~~\rho>a_{p\bar{p}}.
   \end{array} \right.
\end{equation}
where $a_{p\bar{p}}\simeq 2~\mbox{fm}$, $c\simeq 2.161$ , $m_p$ is
the mass of proton and $\lambda=1/(2m_p)\simeq 0.1 {\rm fm}$. In
this case, the Schr${\rm \ddot{o}}$dinger equation for $S$-wave
bound states is
\begin{equation}\label{schrodinger}
{-1\over 2(m_p/2)}{\pa^2\over \pa \rho^2}
u(\rho)+\left[V(\rho)-E\right]u(\rho)=0,
\end{equation}
where $u(\rho)=\rho\;\psi(\rho)$ is the radial wave function, and
$m_p/2$ is the nucleon reduced mass. This equation can be solved
analytically, and has a bound state $u_2(\rho)$  with binding
energy $E_2\simeq -17.2~\mbox{MeV}$  due to the attractive square
well potential at intermediate ranges  (see Appendix). This
molecular state is identified as the  $X(1860)$.

\begin{figure}[hptb]
   \centerline{\psfig{figure=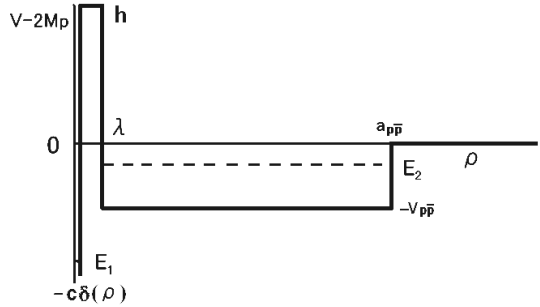,width=7.5cm}}
 \centering
\vskip-0.3in
\begin{minipage}{3.45in}
    \caption{The double well potential of $p\bar{p}$-system.}
\end{minipage}
\end{figure}
\noindent We use this solution to estimate the
$(p\bar{p})$-collision frequency  inside the $X(1860)$. Because of
the effect of $(p\bar{p})$-annihilation, this bound state is not
stable. Note that in this model,
 there are two
attractive potential wells: one is at $\rho\sim 0$ and the other
is at intermediate values; between them is a potential barrier.
 When  $\rho\sim 0$, the
proton and antiproton are in collision.
Therefore, we can derive the $(p\bar{p})$-collision frequency in
the inside of $X(1860)$ by calculating the quantum tunnelling
effect for $u_2(\rho)$ passing through the potential barrier. In
the WKB-approximation, the tunnelling coefficient (i.e., barrier
penetrability) is (see Schiff's book in ref.\cite{Ma})
\begin{eqnarray}\label{WKB}
T_0&=&\exp\left[-2\int_0^{\lambda} dr \sqrt{m_p(h-E_2)}\right]
\nonumber \\  &=&\exp \left[-2\lambda\sqrt{m_p(h-E_2)}\right].
\end{eqnarray}
For the region between  $\rho=\lambda$ to $\rho=a_{p\bar{p}}$, the
time-period $\theta$ for the particle's round trip  is
\begin{equation}
\label{trip} \theta={2\left[a_{p\bar{p}}-\lambda\right]\over
v}=\left[a_{p\bar{p}}-\lambda\right]\sqrt{{m_p\over
V_{p\bar{p}}+E_2}}.~
\end{equation}
Thus, the state $X(1860)$'s (i.e.,$u_2(r)$) lifetime is
$\tau=\theta T_0^{-1}$, and $\nu$,  the the so called
$p\bar{p}$-collision frequency  inside the $X(1860)$, is equal to
$\Gamma_X$,  the total width of $X(1860)$:
\begin{equation}\label{Gamma}
\nu= \Gamma_X\equiv {1\over \tau}= {1\over a_{p\bar{p}}-\lambda}
\sqrt{{ V_{p\bar{p}}+E_2 \over m_p}} \nonumber
 \exp \left[-2\lambda\sqrt{m_p(h-E_2)}\right].~~
\end{equation}
Substituting $E_2=-17.2~\mbox{MeV},\;a_{p\bar{p}}=2.0~\mbox{fm}$
into Eq.(\ref{Gamma}), we obtain the prediction
\begin{equation}\label{GG}
\Gamma_X\simeq 15.5~\mbox{MeV},
\end{equation}
which is compatible with the experimental data \cite{BES}.

The corresponding  $(p\bar{p})$-collision frequency inside the
 $X(1860)$ is about
\begin{equation}\label{nu}
\nu=\Gamma_X/\hbar\simeq 2.35\times 10^{22}~~{\rm Hz}.
\end{equation}
Note that because  the binding energy $E_2$ is rather small
(compared to $2m_p$), the annihilation processes that cause
$X(1860)$ to be unstable occur nearly at rest.

\section{Coherent states with fixed $G-$ and $P-$parities}

When the proton and antiproton collide, they will rapidly
annihilate into mesons. The coherent state method in \cite{batch1}
and \cite{batch2} investigates $(p\bar{p})$-annihilation at rest
without consideration of the $P-$ and $G-$parity. In this paper we
are concerned with annihilation inside the $X(1860)$ where the
$(p\bar{p})$ state has  $I^G(J^{PC})=0^+(0^{-+})$. For this state,
the allowed processes are ($X(1860)\rightarrow \eta 2\pi;~~\eta
4\pi;\cdots \eta 2n \pi;~~3\eta$) (we ignore processes that
involve K-mesons), where the $\pi$ and $\eta$ are both
pseudoscalars and the $G$-parities  are negative and positive
respectively. In this case, we have to introduce $P-$ and
$G-$parities into the previous coherent state description of
mesons radiated in $(p\bar{p})$-annihilations\cite{batch1,batch2}.

We construct coherent states with fixed four momentum and fixed
isospin, and, also, with fixed $G$ parity and $P$ parity.
Following the
 method in \cite{batch1}\cite{batch2}, we first construct the
field operator $F$ that creates a $\pi$ or $\eta$ at the space
time point $x$ and directed in the isospin direction
$\hat{\textbf{n}}$.
\begin{equation}
\label{1.29} F(x,\hat{\textbf{n}})= \int d^3 {\bf k} e^{-ik\cdot
x}f(\textbf{k})\textbf{a}^{\dag}_{\textbf{k}}\cdot\hat{\textbf{n}}
+\int d^3 {\bf q} e^{-iq\cdot x} g(\textbf{q})
b^{\dag}_{\textbf{q}},
\end{equation}
where $k\cdot x=k_0 t-\textbf{k}\cdot\textbf{x}$ with
$k_0=\sqrt{\textbf{k}^2+m^{2}_{\pi}}$,  $q\cdot x=q_0
t-\textbf{q}\cdot\textbf{x}$ with
$q_0=\sqrt{\textbf{q}^2+m^{2}_{\eta}}$,
$\textbf{a}^{\dag}_{\textbf{k}} $ is the isospin-triplet creation
operator, and $b^{\dag}_{\textbf{k}} $ is the isospin-singlet
creation operator.~~From the $G$- an $P$-parities of the $\pi$ and
$\eta$, we have
\begin{eqnarray}
\nonumber \hat{G} {\bf{\pi}}(\textbf{x},t) \hat{G}^{\dag}&=&-{\bf{\pi}}(\textbf{x},t)\\
\nonumber \hat{G} \eta(\textbf{x},t) \hat{G}^{\dag}&=&\eta(\textbf{x},t) \\
\nonumber \hat{P} {\bf{\pi}}(\textbf{x},t) \hat{P}^{\dag}&=&-{\bf{\pi}}(-\textbf{x},t)\\
\label{1.30} \hat{P}
\eta(\textbf{x},t)\hat{P}^{\dag}&=&-\eta(-\textbf{x},t),
\end{eqnarray}
where $\hat{G}$,~$\hat{P}$ are the unitary operators as follows
\begin{eqnarray}
\nonumber
\hat{P}&=& \exp[i\frac{\pi}{2}\sum_{j,\textbf{k}}(a^{+}_{\textbf{k},j}a_{-\textbf{k},j}+b^{+}_{\textbf{k}}b_{-\textbf{k}}+a^{+}_{\textbf{k},j}a_{\textbf{k},j}+b^{+}_{\textbf{k}}b_{\textbf{k}})]\\
\nonumber \hat{G}&=& \exp[i\frac{\pi}{2}\sum_{j,\textbf{k}}(a^{+}_{\textbf{k,-1}}a_{\textbf{k},1}+a^{+}_{\textbf{k,1}}a_{\textbf{k},-1}-a^{+}_{\textbf{k},1}a_{\textbf{k,1}}-a^{+}_{\textbf{k},-1}a_{\textbf{k,-1}})]\times\\
\label{1.31}&&
\exp[-\frac{\pi}{\sqrt{2}}\sum_{\textbf{k}}(a^{+}_{\textbf{k},0}a_{\textbf{k,1}}+a^{+}_{\textbf{k},0}a_{\textbf{k,-1}}-a^{+}_{\textbf{k},1}a_{\textbf{k},0}-a^{+}_{\textbf{k},-1}a_{\textbf{k},0})].
\end{eqnarray}
It is straight-forward to check the following equations
\begin{eqnarray}
\nonumber \hat{G}a^{\dag}_{\textbf{p},i}\hat{G}^{\dag}&=&-a^{\dag}_{\textbf{p},i }\\
\nonumber \hat{G} b^{\dag}_{\textbf{q}}\hat{G}^{\dag}&=&b^{\dag}_{\textbf{q}}\\
\nonumber \hat{P} a^{\dag}_{\textbf{p},i}\hat{P}^{\dag}&=&-a^{\dag}_{-\textbf{p},i }\\
\label{1.32} \hat{G}
b^{\dag}_{\textbf{q}}\hat{G}^{\dag}&=&-b^{\dag}_{-\textbf{q}}
\end{eqnarray}
where $i=1,0,-1$ corresponding to $\pi^{+},\pi^{0},\pi^{-}$. Under
G transformation, $F(x,\hat{\textbf{n}})$ becomes
\begin{equation}
\label{1.33}G(x,\hat{\textbf{n}})=\hat{G} F(x,\hat{\textbf{n}})
\hat{G}^{\dag}=-\int d^3 {\bf k} f(\textbf{k})
\textbf{a}^{\dag}_{\textbf{k}}\cdot\hat{\textbf{n}}~~e^{-ik\cdot
x}+\int d^3 {\bf q} e^{-iq\cdot x} g(\textbf{q})
b^{\dag}_{\textbf{q}}.
\end{equation}
For simplicity we take $f(-\textbf{k})=f(\textbf{k})$ and
$g(-\textbf{q})=g(\textbf{q})$, then the P transformation of
$F(x,\hat{\textbf{n}})$ is as follows
\begin{equation}
\label{1.34} \hat{P} F(x,\hat{\textbf{n}}) \hat{P}^{\dag}=-\int
d^3 {\bf k} f(\textbf{k})
\textbf{a}^{\dag}_{\textbf{k}}\cdot\hat{\textbf{n}}~~e^{-ik\cdot
x^{'}}-\int d^3 {\bf q} e^{-iq\cdot x^{'}}g(\textbf{q})
b^{\dag}_{\textbf{q}}=-F(x^{'},\hat{\textbf{n}}),
\end{equation}
with $\textbf{x}^{'}=(-\textbf{x},t)$. Then the desired coherent
state with fixed four-momentum, fixed isospin, and also with
well-defined G parity(+) and P parity(--) is constructed as
follows
\begin{eqnarray}
\label{1.35}|K,I,I_z\rangle=&\int \frac{d^4x}{(2\pi)^4}
\frac{d\Omega_{\hat{\textbf{n}}}}{\sqrt{4\pi}}~e^{iK\cdot
x}|f,g,x,\hat{\textbf{n}},2\rangle
Y^{*}_{I,I_z}(\hat{\textbf{n}}),
\end{eqnarray}
where
\begin{eqnarray}\label{1.36}
|f,g,x,\hat{\textbf{n}},2\rangle&=&[e^{F(x,\hat{\textbf{n}})}
+e^{G(x,\hat{\textbf{n}})}-F(x,\hat{\textbf{n}})-G(x,\hat{\textbf{n}})
-e^{-F(x^{'},\hat{\textbf{n}})}-e^{-G(x^{'},\hat{\textbf{n}})}\\
\nonumber
&&-F(x^{'},\hat{\textbf{n}})-G(x^{'},\hat{\textbf{n}})]|0\rangle.
\end{eqnarray}
Here we have subtracted the states without a meson  and with only
 one meson, since they violate the conservation of energy
and momentum. The states defined in Eq.(\ref{1.35}) are orthogonal
\begin{equation}
\label{1.37}\langle
K,I,I_z|K^{'},I^{'},I^{'}_z\rangle=\delta^{4}(K-K^{'})\delta_{I
I^{'}}\delta_{I_z I^{'}_z} {\cal I}(K),
\end{equation}
where ${\cal I} (K)$ is the normalization factor:
\begin{eqnarray}
\nonumber&{\cal I}(K)=&\hskip-0.2in \int
\frac{d^{4}x}{(2\pi)^4}\frac{d\Omega_{\hat{\textbf{n}}}d\Omega_{\hat{\textbf{n}}^{'}}}{4\pi}e^{iK\cdot
x}Y^{*}_{I I_z}(\hat{\textbf{n}})Y_{I^{'}
I^{'}_z}(\hat{\textbf{n}}^{'})\{4\exp[\rho_f(x)\hat{\textbf{n}}^{'}\cdot\hat{\textbf{n}}+\rho_{g}(x)]
+4\exp[-\rho_f(x)\hat{\textbf{n}}^{'}\cdot\hat{\textbf{n}}+\rho_{g}(x)]   \\
\label{1.38}&&-8\rho_{g}(x)-4\exp[-\rho_f(x)\hat{\textbf{n}}^{'}\cdot\hat{\textbf{n}}-\rho_{g}(x)]
-4\exp[\rho_f(x)\hat{\textbf{n}}^{'}\cdot\hat{\textbf{n}}-\rho_{g}(x)]-8\rho_{g}(x)\},
\end{eqnarray}
where
\begin{eqnarray}
\nonumber \rho_{f}(x)=\int d^3p|f(\textbf{p})|^2e^{ip\cdot x}\\
\label{1.39} \rho_{g}(x)=\int d^3q|g(\textbf{q})|^2e^{iq\cdot x}.
\end{eqnarray}
We use the expansion method developed in \cite{Yang lu3} to
calculate the normalization integral
\begin{equation}
\label{1.40}{\cal I}(K)=\sum_{m+n \geq 2;\;\; m {\rm \;is\;
even},\; n {\rm \;is\; odd}} \frac{16~I(K,m,n)}{m! n!}F(m,I)
\end{equation}
where
\begin{equation}
\label{1.41}I(K,m,n)= \int
\delta^{4}(K-\sum_{i=1}^{m}p_i-\sum_{j=1}^{n}q_j)\prod_{i=1}^{m}d^3\textbf{p}_i|f(\textbf{p}_i)|^2
\prod_{j=1}^{n}d^3\textbf{q}_j|g(\textbf{q}_j)|^2
\end{equation}
and
\begin{eqnarray}
\nonumber F(m,I)&=&\int
\frac{d\Omega_{\hat{\textbf{n}}}d\Omega_{\hat{\textbf{n}}^{'}}
}{4\pi}Y^{*}_{I
I_z}(\hat{\textbf{n}})Y_{II_z}(\hat{\textbf{n}}^{'})(\hat{\textbf{n}}\cdot\hat{\textbf{n}}^{'*})^{m}\\
\label{1.42}& =&\left\{   \begin{array}{ll}
0 & I >  m \mbox{ or } I-m \mbox{ is odd} \\
\frac{ m! } { (m-I)!! (I+m+1)!! } & I\le m \mbox{ and } I-m \mbox{
is even}.
\end{array}   \right.
\end{eqnarray}
Note that the effect of phase space for the decay  has been taken
into account via the
$\delta^4(K-\sum_{i=1}^{m}p_i-\sum_{j=1}^{n}q_j)\prod_{i=1}^{m}d^3\textbf{p}_i|
\prod_{j=1}^{n}d^3{\bf q_j}$ in function $I(K,m,n)$. Each
individual term with ($m, n$) in the sum Eq.(\ref{1.40})
represents the contribution from the decay channel whose final
particles are $m\pi$ plus $n\eta$. For a fixed total energy, the
sum must terminate. The coherent state naturally gives the result
that only decays to even numbers of $\pi$'s and odd numbers of
$\eta$'s. The  decays conserve $P-$parity and $G-$parity. The mean
numbers of $\pi$ of isospin type $i$ and $\eta $ are given by
\begin{eqnarray}
\nonumber \overline{N}_{{\pi}_i}&=&\frac{1}{{\cal I}(K)}~\langle
K,I,I_{z}|\int
d^3 {\bf k} ~~a^{+}_{\textbf{k},i}a_{\textbf{k},i}|K,I,I_{z}\rangle  \\
\nonumber&=&\frac{1}{{\cal
I}(K)}\int\frac{d^4x}{(2\pi)^4}\frac{d\Omega_{\hat{\textbf{n}}}d\Omega_{\hat{\textbf{n}}^{'}}}{4\pi}e^{iK\cdot
x}Y^{*}_{I,I_z}(\hat{\textbf{n}})Y_{I,I_z}(\hat{\textbf{n}}^{'})\left(
4\rho_{f}(x)
\hat{\textbf{n}}^{'*}_i\hat{\textbf{n}}_i\{\exp[\rho_f(x)\hat{\textbf{n}}^{'*}\cdot\hat{\textbf{n}}+\rho_g(x)] \right. \\
\nonumber&&-\exp[-\rho_f(x)\hat{\textbf{n}}^{'*}\cdot\hat{\textbf{n}}+\rho_g(x)]\}
+4\rho_{f}(x)\hat{\textbf{n}}^{'*}_i\hat{\textbf{n}}_i
\{\exp[-\rho_f(x)\hat{\textbf{n}}^{'*}\cdot\hat{\textbf{n}}-\rho_g(x)]\\
\label{1.43}&&-\left.\exp[\rho_f(x)\hat{\textbf{n}}^{'*}\cdot\hat{\textbf{n}}-\rho_g(x)]\}\right)
\end{eqnarray}
and
\begin{eqnarray}
\nonumber \overline{N}_{\eta}&=&~\frac{1}{{\cal I}(K)}\langle
K,I,I_{z}|\int
d^3 {\bf k} ~~b^{+}_{\textbf{k}}b_{\textbf{k}}|K,I,I_{z}\rangle\\
\nonumber&=&\frac{1}{{\cal
I}(K)}\int\frac{d^4x}{(2\pi)^4}\frac{d\Omega_{\hat{\textbf{n}}}d\Omega_{\hat{\textbf{n}}^{'}}}{4\pi}e^{iK\cdot
x}Y^{*}_{I,I_z}(\hat{\textbf{n}})Y_{I,I_z}(\hat{\textbf{n}}^{'})\left(4\rho_g(x)\{\exp[\rho_f(x)
\hat{\textbf{n}}^{'*}\cdot\hat{\textbf{n}}+\rho_g(x)] \right.\\
\nonumber&&+
\exp[-\rho_f(x)\hat{\textbf{n}}^{'*}\cdot\hat{\textbf{n}}+\rho_g(x)]\}
+4\rho_g(x)\{\exp[-\rho_f(x)\hat{\textbf{n}}^{'*}\cdot\hat{\textbf{n}}-\rho_g(x)]\\
\label{1.44}&&+\left.
\exp[\rho_f(x)\hat{\textbf{n}}^{'*}\cdot\hat{\textbf{n}}-\rho_g(x)]\}\right).
\end{eqnarray}
Using the expansion method, we obtain
\begin{equation}
\label{1.45}\overline{N}_{{\pi}_i}=\frac{16}{{\cal
I}(K)}\sum_{\begin{array}{l}
                         m{\rm \;is\; odd}\\
                         n\;{\rm is\; odd}
                         \end{array}}
                         \frac{1}{m!n!}I(K,m+1,n)G_i(m,I,I_z)
\end{equation}
and
\begin{equation}
\label{1.46}\overline{N}_{{\eta}}=\frac{16}{{\cal
I}(K)}\sum_{\begin{array}{l}
                         m{\rm \;is\; even}\\
                         n\;{\rm is\; even}
                         \end{array}}
\frac{1}{m!n!}I(K,m,n+1)F(m,I),
\end{equation}
where
\begin{eqnarray}
\nonumber G_i(m,I,I_z)&=&\int\frac{d\Omega_{\hat{\textbf{n}}}
d\Omega_{\hat{\textbf{n}}^{'}}}{4\pi}Y^{*}_{I,I_z}(\hat{\textbf{n}})Y_{I,I_z}(\hat{\textbf{n}}^{'})
\hat{\textbf{n}}^{'*}_i\hat{\textbf{n}}_i
(\hat{\textbf{n}}\cdot\hat{\textbf{n}}^{'*})^{m}\\
\label{1.47}&=&\sum_{ln}F(m,l)\frac{2l+1}{2I+1}(\langle|l0,10|I0\rangle\langle
ln,1i|II_z\rangle)^2.
\end{eqnarray}

\section{$X(1860)$-decay through $(p-\bar{p})$-annihilation}

Now we  illustrate the $X(1860)$-decays due to
$(p-\bar{p})$-annihilations. In Section II, we have shown how the
 $p$ and $\bar{p}$  meet together in the $X(1860)$ by
using a model that can  be use to interpret the near-threshold
enhancement in the $p-\bar{p}$ mass spectrum in $J/\psi\rightarrow
\gamma p\bar{p}$\cite{BES}. The $p-\bar{p}$-collision (or overlap
) must lead to  the $(p-\bar{p})$-annihilation, and this causes
$X(1860)$-decay. Moreover, in the previous section, the meson
coherent states with fixed $G$- and $P$-parities are constructed
that describe the final meson states radiated by the
$p-\bar{p}$-annihilation. After these preliminaries, the
investigation of $X(1860)$ decay becomes feasible.

 For the case of $X(1860)$, $I=0$ , and from
Eq.(\ref{1.47}) we have
\begin{equation}
\label{1.48}G_1(m,0,0)=G_0(m,0,0)=G_{-1}(m,0,0)=F(m,1)/3.
\end{equation}
Instituting Eq.(\ref{1.48}) into Eq.(\ref{1.45}), we get
\begin{equation}
\label{1.49}\overline{N}_{{\pi}^+}:\overline{N}_{{\pi}^0}:\overline{N}_{{\pi}^{-}}=1:1:1.
\end{equation}
This indicates that among the products of $X(1860)$-meson decays,
the ratios between the number of $\pi^+$ and $\pi^-$, and between
that for $\pi^\pm$ and $\pi^0$ are fixed.

The probability of the decay with annihilation-products of
($N_{\pi}\pi$, $N_{\eta}\eta$) is given by
\begin{eqnarray}
\nonumber
P(N_{\pi},N_{\eta})&=&\frac{1}{N_{\pi}!N_{\eta}!}\int\prod_{i=1}^{N_{\pi}}d^3{\bf
p}_i \prod_{j=1}^{N_{\eta}}d^3{\bf q}_j~ |\langle \textbf{p}_1
\textbf{p}_2 \cdot\cdot\cdot \textbf{p}_{N_{\pi}} \textbf{q}_1
\textbf{q}_2\cdot\cdot\cdot \textbf{q}_{N_{\eta}}|K,I,I_z\rangle|^2\\
\label{1.50}&=&\frac{1}{{\cal I}(K)}\frac{16
I(K,N_{\pi},N_{\eta})F(N_{\pi},I)}{N_{\pi}!N_{\eta}!}.
\end{eqnarray}
In the same spirit, for the case of ($N_{\pi^+}\pi^+,
N_{\pi^-}\pi^-, N_{\pi^0}\pi^0, N_{\eta}\eta$)
annihilation-products , the probability is
\begin{equation}
\label{1.51}P(N_{\pi^{+}},N_{\pi^{0}},N_{\pi^{-}},N_{\eta})=\frac{1}{{\cal
I}(K)}\frac{16
I(K,N_{\pi^{+}}+N_{\pi^{0}}+N_{\pi^{-}},N_{\eta})F(N_{\pi^{+}},N_{\pi^{0}},N_{\pi^{-}},I)}
{N_{\pi^{+}}!N_{\pi^{0}}!N_{\pi^{-}}!N_{\eta}!},
\end{equation}
where
\begin{equation}
\label{1.52}F(N_{\pi^{+}},N_{\pi^{0}},N_{\pi^{-}},I)=\int
\frac{d\Omega_{\hat{\textbf{n}}}d\Omega_{\hat{\textbf{n}}^{'}}
}{4\pi}Y^{*}_{I
I_z}(\hat{\textbf{n}})Y_{II_z}(\hat{\textbf{n}}^{'})(\hat{\textbf{n}}_+\cdot\hat{\textbf{n}}_{+}^{'*})^{N_{\pi^{+}}}
(\hat{\textbf{n}}_0\cdot\hat{\textbf{n}}_{0}^{'*})^{N_{\pi^{0}}}(\hat{\textbf{n}}_-\cdot\hat{\textbf{n}}_{-}^{'*})^{N_{\pi^{-}}}.
\end{equation}
Since the branching fraction $Br_0(X(1860)\rightarrow m\pi+n\eta)$
is proportional to $P(m,n)$, from Eq.(\ref{1.50}) we can obtain
the ratio of $Br_0(X\rightarrow \eta 4\pi)$ and $Br_0(X\rightarrow
\eta 2\pi)$ as
\begin{equation}
\label{1.53} \frac{Br_0(X\rightarrow \eta4\pi)}{Br_0(X\rightarrow
\eta 2\pi)}=\frac{I(K,4,1)F(4,0)}{4!}\frac{2!}{I(K,2,1)F(2,0)}
=\frac{I(K,4,1)}{20~I(K,2,1)}
\end{equation}
and
\begin{equation}
\label{1.54}\frac{Br_0(X\rightarrow \eta2\pi)}{Br_0(X\rightarrow
3\eta )}=\frac{I(K,2,1)F(2,0)}{2!}\frac{3!}{I(K,0,3)F(0,0)}
=\frac{I(K,2,1)}{I(K,0,3)}.
\end{equation}
Note that in the above coherent state calculation of
$Br_0(X(1860)\rightarrow m\pi+n\eta)$,  charge conservation was
not  taken into account. Here we expand our discussion to include
 charge conservation. Using
$Br(X(1860)\rightarrow m\pi+n\eta)$ to denote the corresponding
branch fraction,  we have
\begin{eqnarray}
\nonumber Br(X\rightarrow \eta2\pi)&=&Br(X\rightarrow
\eta\pi^+\pi^-)+Br(X\rightarrow \eta2\pi^0)\\
\label{1.55}Br(X\rightarrow \eta4\pi)&=&Br(X\rightarrow
\eta2\pi^{+}2\pi^{-})+Br(X\rightarrow
\eta\pi^{+}\pi^{-}2\pi^0)+Br(X\rightarrow \eta4\pi^0).
\end{eqnarray}
Consequently, the ratios between these branch fractions are
\begin{eqnarray}
\nonumber \frac{Br(X\rightarrow \eta2\pi)}{Br(X\rightarrow 3\eta
)}&=&\frac{I(K,2,1)}{I(K,0,3)}\frac{2}{3}\\
\label{1.56}\frac{Br(X\rightarrow \eta4\pi)}{Br(X\rightarrow \eta
2\pi)}&=&\frac{I(K,4,1)}{I(K,2,1)}\frac{7}{300}.
\end{eqnarray}
 Since $m_{\pi}\approx 139 \mbox{MeV}$ and $m_{\eta}\approx 547\mbox{MeV}$,
 the $X(1860)$
 can decay to $\eta 2\pi$, $3\eta$, $\eta 4\pi$, $\eta 6
 \pi$, and $\eta 8\pi$, but not into $3\eta 2\pi$ and still
 conserve  energy.
We investigate the decay channels of $X(1860)\rightarrow (\eta
2\pi,~~\eta 4\pi,$ and $3\eta)$ below.

\par
Following ref.\cite{Yang lu3}, we suppose that the meson field
source turns on at $t=0$ and then decays exponentially in time,
and that it has a spherical symmetric Yukawa shape. In this case,
$f({\bf k})$ (as a Fourier transformation of the meson field
source) is
\begin{equation}
\label{1.57}|f(\textbf{k})|^2=\frac{C~
\textbf{k}^2}{(\textbf{k}^2+\alpha
^2)^2(\omega^2_{\textbf{k}}+\gamma^2)^2 \omega^2_{\textbf{k}}},
\end{equation}
where $\omega_{\textbf{k}}=\sqrt{\textbf{k}^2+m_{\pi}^2}$,  C is a
strength and can be fixed by required that the average energy be
the energy released in annihilation, namely $2m_p$. In units of
pion masses $(m_{\pi}=1)$, by \cite{Yang lu3} we take
$\alpha=\gamma=2$. This corresponds to an annihilation region with
a time and distance scale of half a pion Compton wave length---a
reasonable size that gives  reasonable agreement with experimental
data\cite{batch1,batch2}. Since both $\pi$ and $\eta$ belong to
the pseudoscalar meson octet, we expect that $g(\textbf{k})$
should be the same as $f(\textbf{k})$ except which $m_{\pi}$
replaced by $m_{\eta}$. With this parameter choice, without
consider charge conservation,  we can obtain from
Eq.(\ref{1.53})and Eq.(\ref{1.54}):
\begin{eqnarray}
\nonumber \frac{Br_0(X\rightarrow \eta4\pi)}{Br_0(X\rightarrow
\eta
2\pi)}&\approx&3.8\times10^4\\
\label{1.58}\frac{Br_0(X\rightarrow \eta2\pi)}{Br_0(X\rightarrow
3\eta)}&\approx&8.8.
\end{eqnarray}
When charge conservation is taken into account, using
Eq.(\ref{1.56}), the results become as  follows
\begin{eqnarray}
\label{1.60} \frac{Br(X\rightarrow \eta4\pi)}{Br(X\rightarrow \eta
2\pi)}&\approx&1.8\times10^4\\
\label{1.59}\frac{Br(X\rightarrow \eta2\pi)}{Br(X\rightarrow
3\eta)}&\approx&5.9.
\end{eqnarray}
From Eq.(\ref{1.60}) we find out that $Br(X\rightarrow \eta 2\pi)$
is heavily suppressed by about 4 orders of magnitude compared to
$Br(X\rightarrow \eta 4\pi)$. Namely, the most favorable decay
channel is ($X\rightarrow \eta 4\pi)$, rather than $(X\rightarrow
\eta 2\pi)$. This prediction may not be quantitatively exact, but
 must be  qualitatively correct. We expect  this to be  a
significant feature of the decays of $(p\bar{p})$-bound states
with $I^G(J^{PC})=0^+(0^{-+})$ due to the nucleon-antinucleon
annihilation decay mechanism, which is significantly different
from the naive argument for the decays of ordinary particle as
discussed above  in the Introduction. Thus, we conclude that an
experimental check of this prediction  is meaningful for
distinguishing the baryonium interpretation for the near-threshold
enhancement effects in $(p\bar{p})$-mass spectrum in
$J/\psi{\to}\gamma p{\bar p}$ from other possible interpretations.
In the other hand, such experiments should be valuable efforts
also because they will belong  to seek new evidence for the
existence of exotic hadron $X(1860)$.

Equation (\ref{1.59}) shows $Br(X\rightarrow 3\eta)<<
Br(X\rightarrow \eta 2\pi)$. This is mainly due to the effects of
the decay phase space, and the result is reasonable.

\section{Summary and discussion}

One of the possible interpretation of the near-threshold
enhancement  in the $(p\bar{p})$-mass spectrum in
$J/\psi{\to}\gamma p{\bar p}$ is the existence of a narrow
baryonium resonance $X(1860)$. The mesonic decays of the $X(1860)$
due to the nucleon-antinucleon annihilation have been investigated
in this paper. In order to clarify the picture of
proton-antiproton annihilations inside $X(1860)$, we employed a
toy double well potential to derive the $(p\bar{p})$-collision
frequency, (or collision possibility) inside the $X(1860)$. In
this model,  the annihilations cause  $X(1860)$ decays.
Specifically, in the model the proton and the antiproton  are
separated by a potential barrier, and the
$(p-\bar{p})$-collision's frequency is computed  by considering
quantum tunnelling effects. We further construct meson coherent
states with fixed $G$-parity and $P$-parity. These enable us to
extend the Amado-Cannata-Dedoder-Locher-Shao formulation to
discuss the decays of the $(p\bar{p})$-bound state $X(1860)$. In
this formalism, the process of pseudoscalar meson radiation from
the annihilation is rapid, and, hence, the mesons are classical,
and can be approximately described by coherent states. By this
method, the ratios between the $Br(X\rightarrow \eta 4\pi)$,
$Br(X\rightarrow \eta 2\pi)$ and $Br(X\rightarrow 3\eta )$
branching fractions are derived. Taking appropriate meson field
source functions, and evaluating the integrals related to
 3-body and 5-body phase space in the decay processes, we obtain
quantitative predictions. We find that in contrast to naive
arguments,  the  $Br(X\rightarrow \eta 2\pi)$ is heavily
suppressed about four orders of magnitude in comparison to
$Br(X\rightarrow \eta 4\pi)$. In other words, if $X(1860)$ is a
bound state of $(p\bar{p})$, the most favorable decay channel must
be ($X\rightarrow \eta 4\pi)$, rather than $(X\rightarrow \eta
2\pi)$. This provides a criteria for distinguishing the baryonium
interpretation for the near-threshold enhancement effects in
$(p\bar{p})$-mass spectrum in $J/\psi{\to}\gamma p{\bar p}$ from
other possibilities. Experimental checks are needed.

The unexpected result in Eq.(\ref{1.60}) results  from
calculations based the coherent state theory that successfully
describes nucleon-antinucleon annihilations. This can be seen from
an  intuitive picture.  Naively, the number of valence quarks in
$X(1860)$(or $(p\bar{ p})$) is equal to  the number of valence
quarks in $(\eta\pi\pi),$ i.e., in both systems there are three
quarks plus three anti-quarks, so it seems that the decay
$X\rightarrow \eta 2\pi$ should be the most easily accomplished.
However, the gluon content for $(p\bar{p})$ and $(\eta\pi\pi)$ are
different. Generically, the gluon mass-percentage in the proton
(or antiproton) is larger than that for the $\pi$ or $\eta$. This
point can be seen from some QCD-inspired models, for example the
Skyrme model\cite{Skyrme,Witten,Adkin}. In the chiral limit of
this model (i.e., $m_{\rm quarck}=0$), the masses of the baryons,
including the proton, are non-zero, but the masses of $\pi$ and
$\eta_8$ (that is the main component of $\eta$) vanish. Thus, one
could interpret this as being due to more gluons in baryons which
make them massive even in the limit of massless  quarks. This
indicates that there are some "redundant gluons" that are left
over in the process of $X\equiv(p\bar{p})\rightarrow \eta 2\pi$.
Consequently, the process might be expressed as
\begin{eqnarray}\label{end1}
X\equiv(p\bar{p})\rightarrow \eta2\pi G,
\end{eqnarray}
where $G$ represents the "redundant gluons". Most likely,  $G$ and
$\eta$ could combine to form  the meson $\eta'$, in which case
the process (\ref{end1}) becomes
\newpage
\begin{eqnarray*}
X\rightarrow 2\pi (\eta G)=2\pi \eta'
\end{eqnarray*}
\vskip-0.54in
$$\hskip1.15in |$$ \vskip-0.58in \begin{equation}\label{end} \hskip1.74in \longrightarrow
\eta \pi\pi \end{equation} Eq.(\ref{end}) is just $(X\rightarrow
\eta 4\pi)$, where the factor of $\eta'\rightarrow \eta \pi\pi$ is
the dominate channel to be considered (i.e., $Br(\eta'\rightarrow
\eta \pi\pi)\simeq 65\%$)\cite{PDG}. In this view, the process of
($X\rightarrow \eta 2\pi$) will be almost forbidden and the
($X\rightarrow \eta 4\pi$) or ($X\rightarrow \eta' 2\pi$) would be
 most favorable, i.e.,
\begin{eqnarray}\label{end2}
Br(X\rightarrow \eta'2\pi)>> Br(X\rightarrow \eta2\pi).
\end{eqnarray}
Since $m_{\eta'}>>m_\eta$, this is a very unusual result. This can
be tested experimentally.

This analysis could be extended to  a description based on
classical $SU(3)_f$ fields to describe the small $X(1860)$ decay
branching fractions  into K mesons.

\begin{center}
{\bf ACKNOWLEDGMENTS}
\end{center}
We acknowledge Shan Jin  and Bo-Qiang Ma for stimulating
discussions, especially, the discussions with Shan Jin on the
process of $(X\rightarrow \eta' \pi\pi)$  . The authors are also
very grateful to professor S. Olson for his carefully reading the
manuscript of this paper and for his helpful comments. This work
is partially supported by National Natural Science Foundation of
China under Grant Numbers 90403021, and by the PhD Program Funds
of the Education Ministry of China and KJCX2-SW-N10 of the Chinese
Academy.


\begin{appendix}

\section{}


We solve the ${\rm Schr\ddot{o}dinger}$ equation
(\ref{schrodinger}) with the toy model's potential
(\ref{potential}) (see Fig. 1) in the text. Namely, the potential
$V(\rho)$ ( note $\bar{p}$ is at ${\bf r}_{\bar{p}}=0$,  $p$ is at
${\bf r}_p=x{\bf i}+y{\bf j}+z{\bf k}$, and
$\rho=\sqrt{x^2+y^2+z^2}$ ) is
\begin{equation}\label{potentialA}
V(\rho)=2m_p-c\delta(\rho)+V_c(\rho),
\end{equation}
where
\begin{equation}\label{pA1}
V_c(\rho)=\left\{
   \begin{array}{ll}
   h={m_p/ 4}~~~~~~~~~~~~~~~~~~~~~~~~~~~~~~~~~~~~~~~~ &~~~~ 0<\rho<\lambda\\
   -V_{p\bar{p}}=-2V_{pn}=-73~\mbox{MeV}     &~~~~ \lambda<\rho<a_{p\bar{p}}\\
   0 &~~~~~~~~~ \rho>a_{p\bar{p}}
   \end{array} \right.,
\end{equation}
where $\lambda = 1/(2m_p) = 0.1~\mbox{fm}$ ,
$a_{p\bar{p}}=2.0~\mbox{fm}$, and the equation is
\begin{equation}\label{schrodingerA}
{-1\over 2(m_p/2)}{\pa^2\over \pa \rho^2}
u(\rho)+(V(\rho)-E)u(\rho)=0,
\end{equation}
where $u(\rho)=\rho\psi(\rho)$ is the radial wave function,
$m_p/2$ is the reduced mass, and $0\leq \rho <\infty$. Equation
(\ref{schrodingerA}) is  a one dimensional wave equation with both
one-dimensional delta-function potential ($-c\delta(\rho)$) and
square-well potential, and  can be  solved analytically. We are
interested in the bound state solutions. The corresponding wave
function boundary condition is:
\begin{eqnarray}\label{boundary1}
u(\rho\rightarrow \infty)\rightarrow 0.
\end{eqnarray}
A mathematic trick for solving Eq.(\ref{schrodingerA}) is as
follows: We can mathematically extend the variable $\rho$ from the
region $(0\leq \rho <\infty)$ to $(-\infty < \rho <\infty)$ with
setting $V(-\rho)=V(\rho)$. In this way, $-c\delta(\rho)$ in the
$V(\rho)$ becomes regular. We then can  find out the solutions
$u(\rho)$ by standard procedure to solve Eq.(\ref{schrodingerA})
with boundary condition $u(|\rho|\rightarrow \infty)\rightarrow 0$
. Finally, we take $u(\rho\geq 0)$ (noting, in the physics region
$\rho=\sqrt{x^2+y^2+z^2}\;\geq 0$ $\;\;$) as the physical
solutions which satisfy both differential equation
eq.(\ref{schrodingerA}) and the boundary condition
(\ref{boundary1}).

There are two bound states $u_1(\rho)$ and $u_2(\rho)$:
$u_1(\rho)$ with binding energy $E_1<
-V_{p\bar{p}}=-73~\mbox{MeV}$ is due to $-c\delta(\rho)$-function
potential mainly, and $u_2(\rho)$ with binding energy
$E_2>-73~\mbox{MeV}$ is due to the attractive square well
potential at middle range mainly. They are as follows
\begin{equation}\label{solution1A}
u_1(\rho)=\left\{
   \begin{array}{ll}
   A_1(e^{-\sqrt{m_p(h-E_1)}\;\rho}+\alpha_1 e^{\sqrt{m_p(h-E_1)}\;\rho}),~ &~~~~ 0<\rho<\lambda\\
   A_2(e^{-\sqrt{m_p(-V_{p\bar{p}}-E_1)}\;\rho}+\beta_1 e^{\sqrt{m_p(-V_{p\bar{p}}-E_1)}\;\rho}),    &~~~~ \lambda<\rho<a_{p\bar{p}}\\
   A_3 e^{-\sqrt{-m_pE_1}\;\rho}, &~~~~~~~~~ \rho>a_{p\bar{p}}
   \end{array} \right.,
\end{equation}
\begin{equation}\label{solution2A}
u_2(\rho)=\left\{
   \begin{array}{ll}
   B_1(e^{-\sqrt{m_p(h-E_2)}\;\rho}+\alpha_2 e^{\sqrt{m_p(h-E_2)}\;\rho}),~ &~~~~ 0<\rho<\lambda\\
   B_2\sin(\sqrt{m_p(V_{p\bar{p}}+E_2)}\;\rho+\beta_2),    &~~~~ \lambda<\rho<a_{p\bar{p}}\\
   B_3 e^{-\sqrt{-m_pE_2}\;\rho}, &~~~~~~~~~ \rho>a_{p\bar{p}}
   \end{array} \right.,
\end{equation}
where $\alpha_1,\;\alpha_2,\;\beta_1,\;\beta_2,$ and $A_i,\;B_i$
with $i=1,2,3$ are constants. Due to the potential of
$-c\delta(\rho)$ in the $V(\rho)$, there are relations between $c$
and $\alpha_1,\;\alpha_2$ as follows
\begin{equation}\label{relationA}
c={2(1-\alpha_2)\over 1+\alpha_2}\sqrt{{h-E_2 \over m_p}}
={2(1-\alpha_1)\over 1+\alpha_1}\sqrt{{h-E_1 \over m_p}}.
\end{equation}

We return  to solving the quantum mechanics problem given above.
For $u_1(\rho)$ or $u_2(\rho)$, the wave-function continuum
conditions are as follows
\begin{equation}\label{continue1}
{d\over d\rho}\log u_i(\rho)|_{(\rho=\lambda^-)}= {d\over
d\rho}\log u_i(\rho)|_{(\rho=\lambda^+)},
\end{equation}
\begin{equation}\label{continue2}
{d\over d\rho}\log u_i(\rho)|_{(\rho=a_{p\bar{p}}^-)}= {d\over
d\rho}\log u_i(\rho)|_{(\rho=a_{p\bar{p}}^+)},
\end{equation}
where $i=1,~2$. Then we have
\begin{eqnarray}
\label{eq1}
&~&\sqrt{m_p(h-E_1)}\;{(-e^{-\sqrt{m_p(h-E_1)}\;\lambda}+\alpha_1
e^{\sqrt{m_p(h-E_1)}\;\lambda})\over {
(e^{-\sqrt{m_p(h-E_1)}\;\lambda}+\alpha_1
e^{\sqrt{m_p(h-E_1)}\;\lambda})}}\;
\nonumber \\
&=&\sqrt{m_p(-V_{p\bar{p}}-E_1)}\;{(-e^{-\sqrt{m_p(-V_{p\bar{p}}-E_1)}\;\lambda}+\beta_1
e^{\sqrt{m_p(-V_{p\bar{p}}-E_1)}\;\lambda}) \over
{(e^{-\sqrt{m_p(-V_{p\bar{p}}-E_1)}\;\lambda}+\beta_1
e^{\sqrt{m_p(-V_{p\bar{p}}-E_1)}\;\lambda})}}\;,\\
\label{eq2}
&~&\sqrt{m_p(-V_{p\bar{p}}-E_1)}\;{(-e^{-\sqrt{m_p(-V_{p\bar{p}}-E_1)}\;a_{p\bar{p}}}+\beta_1
e^{\sqrt{m_p(-V_{p\bar{p}}-E_1)}\;a_{p\bar{p}}}) \over
{(e^{-\sqrt{m_p(-V_{p\bar{p}}-E_1)}\;a_{p\bar{p}}}+\beta_1
e^{\sqrt{m_p(-V_{p\bar{p}}-E_1)}\;a_{p\bar{p}}})}}\;
\nonumber \\
&=&-\sqrt{-m_pE_1}\;,\\
\label{eq3}
&~&\sqrt{m_p(h-E_2)}\;{(-e^{-\sqrt{m_p(h-E_2)}\;\lambda}+\alpha_2
e^{\sqrt{m_p(h-E_2)}\;\lambda})\over {
(e^{-\sqrt{m_p(h-E_2)}\;\lambda}+\alpha_2
e^{\sqrt{m_p(h-E_2)}\;\lambda})}}\;
\nonumber \\
&=&\sqrt{m_p(V_{p\bar{p}}+E_2)}\cot\left(\sqrt{m_p(V_{p\bar{p}}+E_2)}\;\lambda+\beta_2\right)\;,\\
\label{eq4}
&~&\sqrt{m_p(V_{p\bar{p}}+E_2)}\cot\left(\sqrt{m_p(V_{p\bar{p}}+E_2)}\;a_{p\bar{p}}+\beta_2\right)\;
\nonumber \\
&=& -\sqrt{-m_pE_2}\;.
\end{eqnarray}
These equations can be solved numerically. $E_1$ is an input of
the model. Following ref.\cite{Yan} and taking $E_1\simeq -976{\rm
MeV}<< -73 {\rm MeV}$ as input (see below), we get the parameters
in the solutions as follows
\begin{eqnarray}\label{solutionA}
\alpha_1&=&0.025,~~~\alpha_2=-0.352, \\ \nonumber \beta_1 &\simeq&
0, ~~~\beta_2=-3.383,\\ \nonumber c&=&2.161,
\end{eqnarray}
from which we get the binding energy of the state $u_2(\rho)$,
\begin{equation}
E_2=-17.2~\mbox{MeV}.
\end{equation}
This is the result that is used in the text.

The functions $u_1(\rho)$ and $u_2(\rho)$ are shown in Fig.2 and
Fig.3 respectively. From the figures, one can see that $u_1(\rho)$
is a sharply peaked curve with a maximum at $\rho=0$, and
$u_2(\rho)$ has a node at $\rho\sim 0.3{\rm fm}$ and $|u_2(\rho)|$
has an absolute maximum at $\rho\sim 1.5{\rm fm}$. Both
$u_1(\rho)$ and $u_2(\rho)$ satisfy the boundary condition given
by Eq.(\ref{boundary1}).
\begin{figure}[hptb]
\begin{center}
\includegraphics*[13pt,13pt][280pt,227pt]{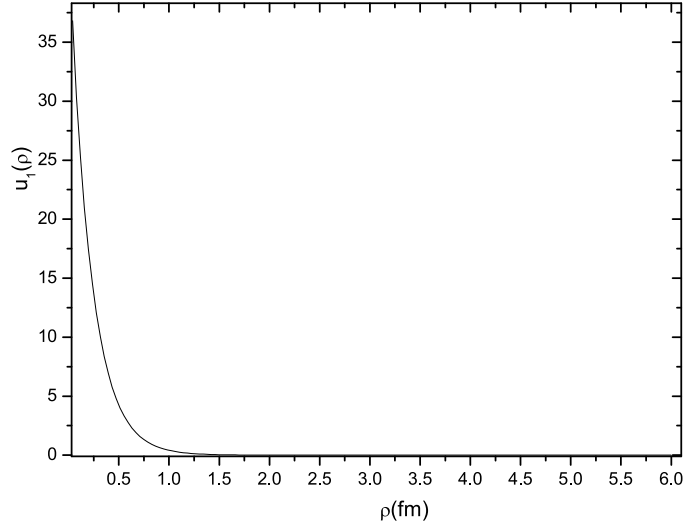}
\caption{The Wave Function $u_1(\rho)$.}
\end{center}
\end{figure}
\begin{figure}[hptb]
\begin{center}
\includegraphics*[13pt,13pt][280pt,227pt]{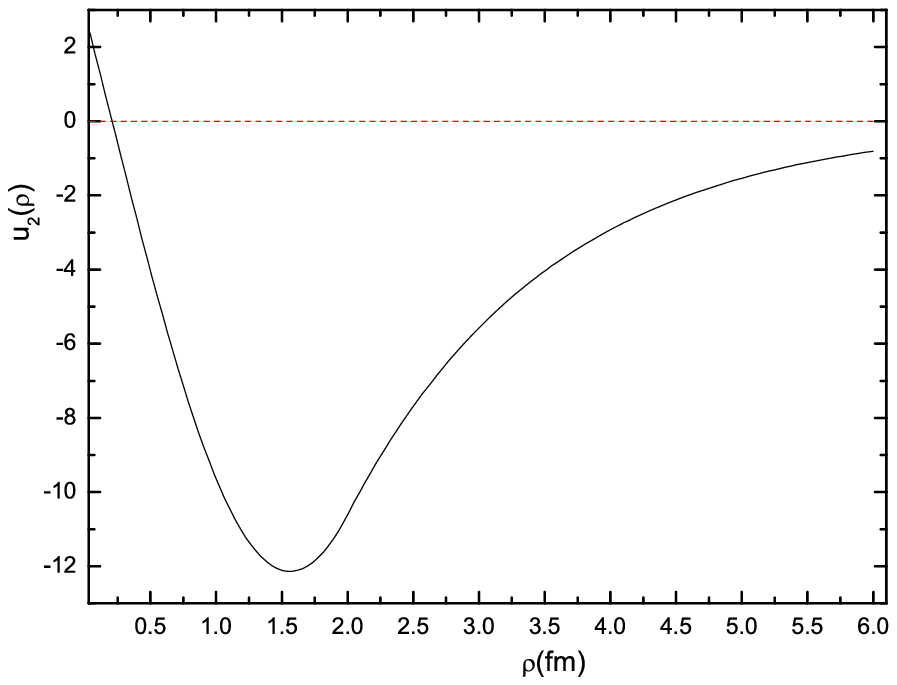}
\caption{The Wave Function $u_2(\rho)$.}
\end{center}
\end{figure}

In order to be sure of that the three-dimensional wave functions
of our solutions  can be normalized, we verify that
\begin{eqnarray}\label{norm}
\int d^3{\bf x} |\psi_i({\bf x})|^2&=&\int d\Omega \int_0^\infty
\rho^2 d\rho {1\over \rho^2}|u_i(\rho)|^2= \int d\Omega
\int_0^\infty  d\rho |u_i(\rho)|^2 \\ \nonumber &=& 4\pi \left(
\int_0^{a_{p\bar{p}}}d\rho |u_i(\rho)|^2 +
\int_{a_{p\bar{p}}}^\infty d\rho |u_i(\rho)|^2 \right)\\
\nonumber &=& 4\pi \left( {\rm const.} + {\rm const.}
\int_{a_{p\bar{p}}}^\infty d\rho e^{-2\sqrt{-m_pE_i}} \right) \\
\nonumber &=& {\rm finite},
\end{eqnarray}
where $i=1,2$ and $u_i(\rho)$ have been given in
eqs.(\ref{solution1A}), (\ref{solution2A}). Equation (\ref{norm})
is a check to the rationality of our solutions.

In the following, we discuss the the parameters in the model in
order:

\begin{enumerate}

\item {\it $a_{p\bar{p}}$ and $V_{p\overline{p}}$}~ : We take the
width of the square well potential, denoted as $a_{p\bar{p}}$, as
close to that of the deuteron,  i.e., $a_{p\bar{p}}\sim
a_{pn}\simeq 2.0 ~\mbox{fm}$. According to QCD inspired
considerations \cite{Datta,Maltman,Rujula}, the well potential
between $q$ and $\bar{q}$ should be twice as attractive as the
$q-q$-case, i.e., the depth of the $p\bar{p}$ square well
potential is $V_{p\overline{p}}\simeq 2V_{pn}=73~\mbox{MeV}$.

\item {\it $h$ and $\lambda$}~: The quantitative results of the
model somehow depend on the parameters of the  potential, such as
the barrier height $2 m_p+h$ and width $\lambda$. For
definiteness, we have taken them to be about $2m_p+m_p/4$ and
$1/(2m_p)$ in solving the Schr$\ddot{o}$dinger equation. Actually,
this is reasonable because the dependence on height and width are
weak in the practice calculation. The results with several
different values for height and width are listed in Tables 1 and
2, where we see that the above values can give a reasonable
binding energy $E_2$ and decay width $\Gamma_X$, in compatible
with the BES measurement.

\vskip0.2in
\begin{center}
\begin{tabular}{|c|c|c|c|c|c|}\hline
barrier height $2m_p+h$ & $2m_p+m_p/8$ & $2m_p+m_p/6$&
$2m_p+m_p/4$ & $2m_p+m_p/2$ & $2m_p+m_p$
\\\hline $E_2$ & $-17.2~\mbox{MeV}$ & $-17.2~\mbox{MeV}$ &
$-17.2~\mbox{MeV}$ & $-17.3~\mbox{MeV}$ & $-17.4~\mbox{MeV}$
\\\hline
 $\Gamma_X$ &  $17.7~\mbox{MeV}$ &  $16.8~\mbox{MeV}$ & $15.5~\mbox{MeV}$ & $12.7~\mbox{MeV}$ &
 $9.7~\mbox{MeV}$  \\\hline
\end{tabular}
\vskip0.1in
\begin{minipage}{4in}
    TABLE 1: The binding energy $E_2$ and width $\Gamma$
    obtained by solving the Skyrmion-type potential model with the potential barrier height $2m_p+h$ from
     $2m_p+m_p/8$ to
    $2m_p+m_p$.
\end{minipage}
\end{center}
\vskip0.2in

\vskip0.2in
\begin{center}
\begin{tabular}{|c|c|c|c|c|c|}\hline
barrier width $\lambda$ & $0.5/(2m_p)$ & $1/(2m_p)$ & $2/(2m_p)$&
$3/(2m_p)$ & $4/(2m_p)$
\\\hline $E_2$ & $-17.1~\mbox{MeV}$ & $-17.2~\mbox{MeV}$ &
$-17.4~\mbox{MeV}$ & $-17.4~\mbox{MeV}$ & $-16.9~\mbox{MeV}$
\\\hline
 $\Gamma_X$ &  $19.3~\mbox{MeV}$ &  $15.5~\mbox{MeV}$ & $9.9~\mbox{MeV}$ & $6.4~\mbox{MeV}$ &
 $4.2~\mbox{MeV}$  \\\hline
\end{tabular}
\vskip0.1in
\begin{minipage}{4in}
    TABLE 2: The binding energy $E_2$ and width $\Gamma$
    obtained by solving the Skyrmion-type potential model with the potential barrier width $\lambda$ from $0.5/(2m_p)$ to
    $4/(2m_p)$ with $h=m_p/4$ fixed.
\end{minipage}
\end{center}

\item {\it $c$ and $E_1$}~: At $\rho\sim 0$, $V(\rho)\sim
-c\;\delta (\rho)$ with a constant $c>0$, which is a free
parameter in the model. $E_1$ is the eigenvalue of $u_1(\rho)$,
which is roughly the energy level of the ${\rm
Schr\ddot{o}dinger}$ equation with the one dimensional
delta-function potential $V(\rho)\sim -c\;\delta (\rho)$. So,
$E_1$ is $c$-dependent and, hence, once $E_1$ were fixed, the free
parameter $c$ fixed. For definiteness, in this paper, we have
taken
\begin{equation}\label{E1} E_1=-(2m_p-m_{\eta_0})\simeq
-976~\mbox{MeV},
\end{equation}
and then the corresponding $c$ value is $c\simeq 2.161$ (see
Eq.(\ref{solutionA})).
\end{enumerate}

\end{appendix}

\end{document}